# One-dimensional Photonic Crystal Structure Enhanced External-Magnetic-Field-Free Spintronic Terahertz High-Field Emitter

Zehao Yang[1,5], Jiahui Li[3,4], Shaojie Liu[2], Zejun Ren[1], Mingxuan Zhang[1], Chunyan Geng[1], Xiufeng Han[3,4], Caihua Wan[3*], and Xiaojun Wu[1,5,6*]

[1]School of Electronic and Information Engineering, Beihang University, Beijing 100191, China.
[2]School of Cyber Science and Technology, Beihang University, Beijing 100191, China.
[3]Beijing National Laboratory for Condensed Matter Physics, Institute of Physics, Chinese Academy of Sciences, Beijing 100190, China.
[4]Center of Materials Science and Optoelectronics Engineering, University of Chinese Academy of Sciences, Beijing 100049, China.
[5]Zhangjiang Laboratory, 100 Haike Road, Shanghai 201210, China.
[6]Wuhan National Laboratory for Optoelectronics, Huazhong University of Science and Technology, Wuhan 430074, China.

**Abstract**
Intense terahertz (THz) radiation in free space offers multifaceted capabilities for accelerating electron, understanding the mesoscale architecture in (bio)materials, elementary excitation and so on. However, the lack of high-beam-quality, high-stability, low-cost and easy-to-implement high-field THz sources is impeding the proliferation of strong-field THz applications. Recently popularized spintronic THz emitters (STEs) with their versatility such as ultra-broadband, cost-effectiveness, large-size and ease-for-integration have become one of the most promising alternative for the next generation of intense THz sources. Nevertheless, the typical $W|Co_{20}Fe_{20}B_{60}|Pt$ necessitates an external magnetic field to saturate magnetization for stable operation, limiting its scalability for achieving higher THz field with uniform distribution over larger sample areas. Here we demonstrate the methodologies of enhancing the high-field THz radiation of external-magnetic-field-free antiferromagnetic metal | ferromagnetic metal | heavy metal ($IrMn_3|Co_{20}Fe_{20}B_{60}|W$) heterostructure via optimizing the substrate with superior thermal conductivity and integrating a one-dimensional photonic crystal (PC) structure to maximize the radiation efficiency. Besides, the integration of the PC featuring high reflectivity enables significant potential in mitigating the semiconductor photoconductive effect. Under the excitation of a Ti: sapphire femtosecond laser amplifier with central wavelength of 800 nm, pulse duration of 35 fs, and repetition rate of 1 kHz and maximum single pulse energy of 5.5 mJ, we successfully generate intense THz radiation with focal peak electric field up to 1.1 MV/cm with frequency range covering 0.1-10 THz without external magnetic fields. These high-field STEs will also enable other applications such as ultrabroadband high-field THz spectroscopy and polarization-based large-size strong-field THz imaging.

# 1 Introduction

High-field terahertz (THz) electromagnetic pulses offer immense potential in all-optical electron acceleration[1-4], non-equilibrium matter state control[5-8], and biological effects[9, 10]. THz waves have natural advantages in driving electron acceleration, as this band can support a higher order of acceleration gradient with smaller corresponding sizes of the acceleration structures[2]. Currently, THz-driven electron accelerators have achieved an energy gain of 1.1 MeV, and the effective acceleration gradient can reach 210 MV/m[3]. Besides, the interaction between high-field THz radiation and matter can directly drive the elementary excitations in solids through resonance excitation, including coherent control of ionic motion[11], lattice oscillations[12], electronic properties[13], molecular orientation and rotation[14]. High-field THz can also non-resonantly induce phenomena that demand significantly higher photon energy, such as field ionization of impurities and excitons[15], high-field electron transport[16], formation of ionization and exciton[17], and impact on material phase transitions[18]. Furthermore, high-field THz pulses have already been applied on biological effect regulation. Several biomaterials, exemplified by short dsDNA illuminated by intense THz radiation with a peak electric field of ~70 kV/cm, reveal the rapid dissociation of short dsDNA at room temperature[9]. Additionally, when the electric field is >0.5 MV/cm, THz pulses can change the local intracellular concentration of $Zn^{2+}$, thereby regulating the arrangement of the cytoskeleton and promoting the adhesion of human induced pluripotent stem cells[10]. However, all the applications mentioned above require high-performance strong-field THz sources.

In recent years, with the rapid advancement of ultrashort femtosecond laser technology, various nonlinear processes have provided various effective methods for producing high-field THz pulses. These methods commonly involve femtosecond laser-driven photoconductive antennas[19, 20], nonlinear optical crystals[21-26], laser-induced plasma in gases[27], liquids[28], and solids[29], and difference frequency generation[30]. Most table-top intense THz sources currently rely on the optical rectification in organic crystals and lithium niobate. The disadvantages of organic crystal-based THz sources lie in the crystal size limitation, prone to deliquescence, low damage threshold, and cannot guarantee their stability[31]. Tilted pulse front technique enabled lithium niobate THz sources have the capability of generating a peak field exceeding 1 MV/cm under the excitation of normal kHz-repetition rate femtosecond amplifiers[24]. However, The optical setup of this technology is relatively complex, requiring the collaborative optimization of laser parameters, and the generated spectrum is not very broad, limiting its application scenarios[23]. If a low-cost, ultra-broadband, large-area, polarization-tunable strong-field THz light source capable of complementing the lithium niobate THz strong source can be identified, it is anticipated that more valuable application potentials of strong-field THz radiation would be unlocked.

The recently emerged spintronic THz emitters (STEs) precisely meet the above requirements[32-35]. Various materials and their corresponding structures, exemplified by $W|Co_{20}Fe_{20}B_{60}|Pt$ trilayer heterostructures have been demonstrated the capabilities of generation ultrabroadband and strong-field THz pulses (>1.0 MV/cm) via inverse spin Hall effect[36, 37]. However, this conventional STEs require an external magnetic field to achieve saturation magnetization, significantly hinders the adoption of larger sizes and higher powers to generate THz radiation with higher electric field and more uniform radiation field distribution. Although the 4-inch antiferromagnetic (AFM) | ferromagnetic (FM) | heavy metal (HM) $IrMn_3|Co_{20}Fe_{60}B_{20}|W$ heterostructure has achieved strong-field THz radiation without an external magnetic field, its focal electric field is <250 kV/cm and cannot be compared with traditional STEs[38]. The main reasons lie in the insufficient utilization of the pump laser and the inadequate thermal management of the substrate.

To address these issues, we not only replaced the substrate material of the $IrMn_3|Co_{20}Fe_{60}B_{20}|W$ sample from $SiO_2$ to high-resistance silicon (Si) wafers, but also fabricated a one-dimensional photonic crystal structure on the substrate that can almost fully utilize the pump laser before depositing into the substrate, raising the focal THz radiation field to 1.0 MV/cm. The benefit of using high-resistance Si as the substrate is that it offers a higher thermal conductivity, which is beneficial for suppressing the thermal accumulation of the emitter, allowing the working temperature of the nanofilm to be lower than the Curie temperature, increasing the emission efficiency saturation threshold of the ultrashort and ultra-intense femtosecond laser pumped heterostructure sample, and thereby enabling stronger THz radiation to be generated per unit area. The one-dimensional photonic crystal not only enables the pump laser to be almost completely utilized to generate terahertz radiation, but also suppresses the issue of reduced terahertz outcoupling efficiency caused by the generation of free carriers in the silicon substrate by the excess pump laser. Through these effective approaches, we have obtained ultrashort THz pulses in 4-inch $IrMn_3|Co_{20}Fe_{60}B_{20}|W$ heterostructure with a pulse duration of 110 fs, a single pulse energy of 62.5 nJ under the excitation pump energy of 5.5 mJ and very uniform radiation field distribution without external magnetic fields.

## 2 Samples and Methods

The structure of the $IrMn_3|Co_{20}Fe_{60}B_{20}|W$ sample employed in this experiment is presented in the lower inset of Figure 1(a). It comprises a 2-nm AFM material ($IrMn_3$), a 2-nm FM material ($Co_{20}Fe_{60}B_{20}$), and a 2-nm HM material (W). These three materials are deposited on the surface of a $SiO_2$ substrate via magnetron sputtering to form a trilayer stacked spintronic heterostructure. Contrary to conventional STEs that depend on an external magnetic field to achieve saturation magnetization in the $Co_{20}Fe_{60}B_{20}$ layer, this sample attains saturation magnetization by pinning the in-plane magnetic field of the $Co_{20}Fe_{60}B_{20}$ layer through exchange bias or coupling effects at the AFM|FM interface. The upper inset of Figure 1(a) depicts the principle of AFM-STEs radiating THz pulses. An incident femtosecond laser prompts the longitudinal spin currents ($\vec{J}_{s\text{-}IrMn_3}$ and $\vec{J}_{s\text{-}W}$) in the $Co_{20}Fe_{60}B_{20}$ layer to flow respectively into the $IrMn_3$ layer and the W layer. Through the inverse spin Hall effect, the longitudinal spin currents are transformed into in-plane charge currents ($\vec{J}_{c\text{-}IrMn_3}$ and $\vec{J}_{c\text{-}W}$), resulting in the radiation of the THz electromagnetic pulse. Taking advantage of the opposite spin Hall angles of IrMn$_3$ and W, spin currents in opposite directions can be converted into charge currents in the same direction, generating in-phase THz pulses in the IrMn$_3$ layer and the W layer. As a consequence, the radiated THz pulses can be amplified and effectively strengthened via coherent superposition. The intensity of THz radiation from AFM-STEs can be acquired through the formula of the inverse spin Hall effect.

$$\vec{E}_{THz} \propto \vec{J}_c = \vec{J}_{c\text{-IrMn}_3} + \vec{J}_{c\text{-W}} = \gamma_{IrMn_3} \cdot \left(\vec{J}_{s\text{-IrMn}_3} \times \frac{\vec{M}}{\vec{M}_s}\right) + \gamma_{W} \cdot \left(\vec{J}_{s\text{-W}} \times \frac{\vec{M}}{\vec{M}_s}\right)$$

where $\vec{M}$ is the magnetic field of the Co$_{20}$Fe$_{60}$B$_{20}$ layer, and $\vec{M}_s$ is the saturated magnetic field of Co$_{20}$Fe$_{60}$B$_{20}$.

In contrast to traditional STEs, AFM-STEs provide more convenient control over the polarization of THz radiation. Due to its electric dipole emission mechanism, there are two methods for directly manipulating the polarization of THz pulses. Firstly, by keeping the laser incidence plane fixed, the polarization of THz can be modified by rotating the sample azimuth. In the experiment, measuring the peak-to-peak change of THz radiation by rotating the sample 360° disclosed the dependence of polarization on the azimuth, as shown in Figure 1(b). The graph presents a distinct bilaterally symmetric distribution resembling an "8" shape, with the peak-to-peak value approaching zero at 90°/270°, indicating the minimum remanence in the hard axis of the Co$_{20}$Fe$_{60}$B$_{20}$ layer. Furthermore, when the pinning direction is horizontal, the THz polarization can be adjusted by flipping the sample surface, as illustrated in Figure 1(c). Flipping the sample horizontally while changing the laser incidence plane does not affect the THz radiation polarization. During the sample flipping, both the magnetic field and spin current directions change simultaneously, maintaining the initial state. The amplitude change is solely caused by the different sequence of the laser and THz passing through the heterostructure and the substrate. However, vertical flipping of the sample disrupts the symmetry of the THz radiation. This is because only the direction of the spin current changes during the flipping process, while the magnetic field direction remains unchanged, leading to a change in the polarization change. As depicted in Figure 1(d), the time-domain signals obtained from both flipping cases display a symmetrical distribution, confirming that the radiation mechanism of AFM-STEs is the electric dipole mechanism.

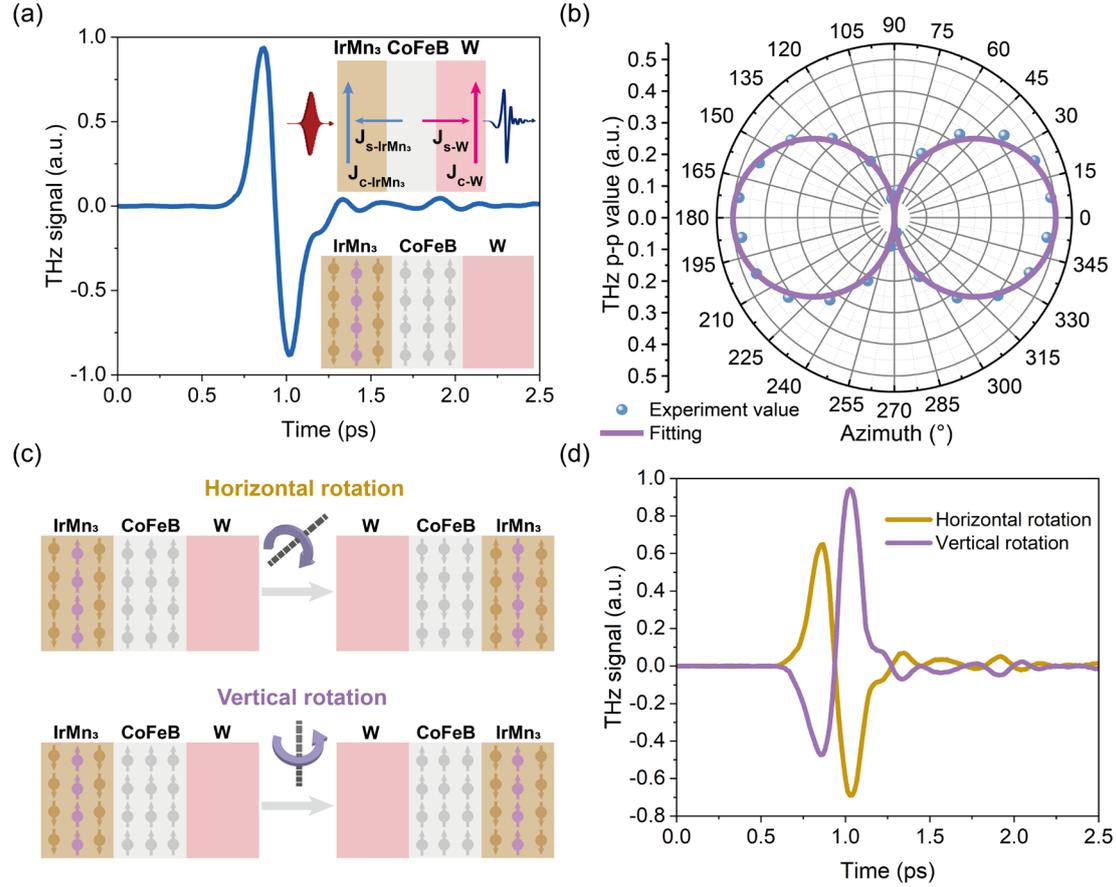

Figure 1. THz radiation from a 4-inch IrMn$_3$-Co$_{20}$Fe$_{60}$B$_{20}$-W heterostructure without an external magnetic field. (a) The THz time-domain signal of the IrMn$_3$-Co$_{20}$Fe$_{60}$B$_{20}$-W heterostructure, the principle of radiating THz (upper inset), and the magnetic field distribution generated by exchange bias or coupling effects (lower inset). (b) The radiation characteristics by rotating the sample azimuth. (c) The schematic diagram of horizontally flipping and vertically flipping the sample when the in-plane magnetic field is horizontally pinned. (d) The THz time-domain signals when the sample is flipped horizontally and vertically.

Similar to conventional STEs, the emission of AFM-STEs is also profoundly affected by the operating temperature. The SiO$_2$ substrates possess poor thermal conductivity, and the laser energy not absorbed by the IrMn$_3$|Co$_{20}$Fe$_{60}$B$_{20}$|W heterostructure is transferred to the substrate, causing an increase in the working temperature. When the transient electronic temperature exceeds the Curie temperature of Co$_{20}$Fe$_{60}$B$_{20}$, the pump energy saturation of AFM-STEs will occur prematurely. To alleviate the influence of substrate heat deposition on the AFM-STEs, we replaced the SiO$_2$ (1.5 W/mK) with Si (150 W/mK) because of its superior thermal conductivity at 300 K. Nevertheless, Si is not an optimal substrate option as it is a photoconductive semiconductor. When exposed to a near-infrared laser on its surface, Si will generate electron-hole pairs, increasing the conductivity within the diffusion range and creating a plasma region[39]. The plasma region is highly degenerate, exhibiting quasi-metallic high conductivity properties, and hinders the transmission of incident THz radiation by reflecting it, significantly hampering the forward radiation[40]. Moreover, the IrMn$_3$|Co$_{20}$Fe$_{60}$B$_{20}$|W heterostructure can only absorb 50%-60% of the pump pulse, and the surplus energy

not only acts as a trigger for the plasmatic state of the Si substrate but also reduces the efficiency of the pump energy

To tackle the aforementioned problems, we incorporated a one-dimensional photonic crystal (PC) layer between the heterostructure and the Si substrate, establishing the Si-PC-AFM-STEs structure, as depicted in Figure 2(a). Figure 2(c) illustrates the composition and operational principle of the PC layer, which is composed of $[HfO_2(92\ nm)|SiO_2(136\ nm)]_x$. By employing the principle of destructive interference, the transmission of laser light is minimized to form a high-reflection layer. This boosts the efficiency of pump pulse absorption by the heterostructure and prevents the photoconductive effect of the Si substrate. The simulation of various PC layer pairs was carried out using the ray optics module in the Comsol Multiphysics software, and the results are presented in Figure 2(b). As the number of PC layer pairs increases, the reflectivity of the PC layer gradually rises, along with a decrease in the spectrum. To diminish the absorption of excess pump laser by the Si substrate, it is essential to minimize the transmittance of the PC layer. In order to meet the processing requirements, we fabricated a PC layer consisting of 40 pairs of $HfO_2(92\ nm)|SiO_2(136\ nm)$ with a total thickness of approximately 9 μm. Figure 2(b) presents the measured results, indicating that the reflectivity of the PC layer exceeds 99.2% in the 750-1050 nm range. The minor deviation between the experiment and simulation results is mainly attributed to the differences in the actual refractive index, film thickness of the material, and simulation parameters during the sputtering process. Figure 2(d) offers a schematic illustration of the Si substrate with and without the PC layer. The high reflectivity of the PC layer in the visible red and near-infrared bands causes the coated Si substrate to appear distinctly magenta compared to the pure Si substrate. Moreover, the THz radiation of STEs are not influenced by the wavelength and polarization of the pump laser. Combined with the broadband characteristics of the PC layer, the operational spectrum of AFM-STEs can cover the typical central wavelengths of the Ti: sapphire femtosecond laser (800 nm) and the fiber femtosecond laser (1030 nm), providing broader application prospects.

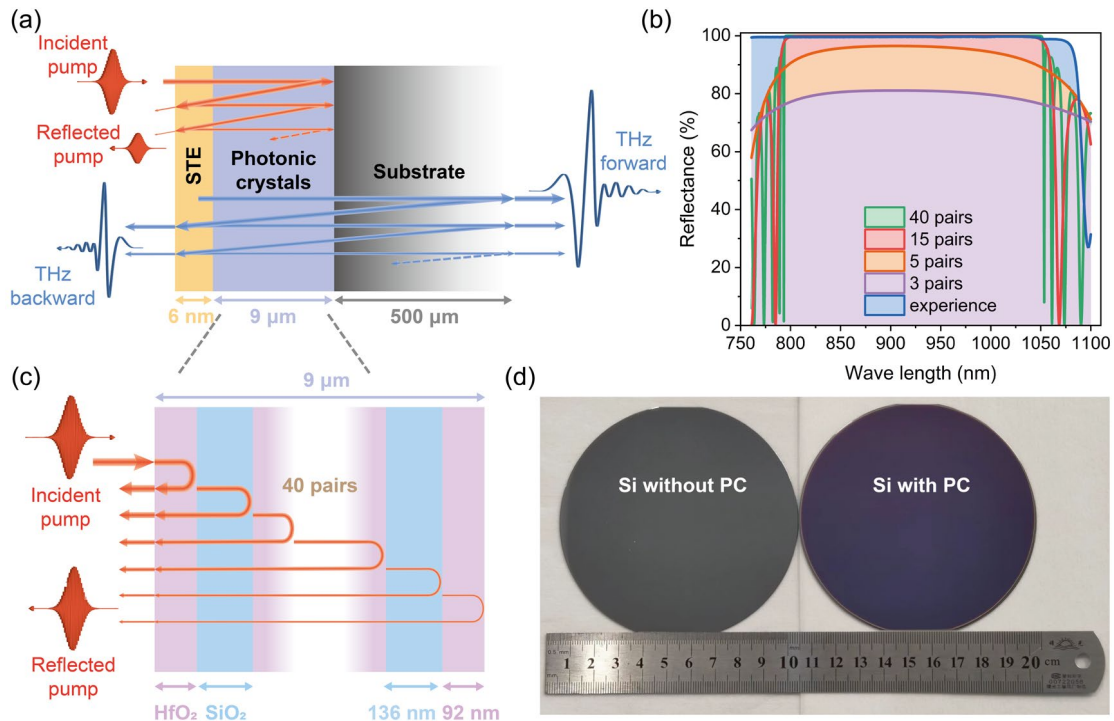

Figure 2. The optimization of AFM-STEs in terms of working temperature and photon. (a) The schematic diagram of the THz radiation principle of Si-PC-AFM-STEs. (b) The reflectivity of different pairs of PC layers in simulation and fabricated PC layers. (c) The schematic diagram of the working principle of the PC layer. (d) The sample diagram of Si substrate with and without PC layer.

We utilized the experimental setup shown in Figure 3 to conduct experiments on the Si-AFM-STEs and Si-PC-AFM-STEs samples to verify the theory mentioned above. A Ti: sapphire laser amplifier with a central wavelength of 800 nm was adopted as the source. It was divided into pump pulse and probe pulse by a beam splitter. A 3× telescope expansion was applied to enlarge the pump spot to 20 mm for the experiment, and traditional electro-optical sampling was used for detection.

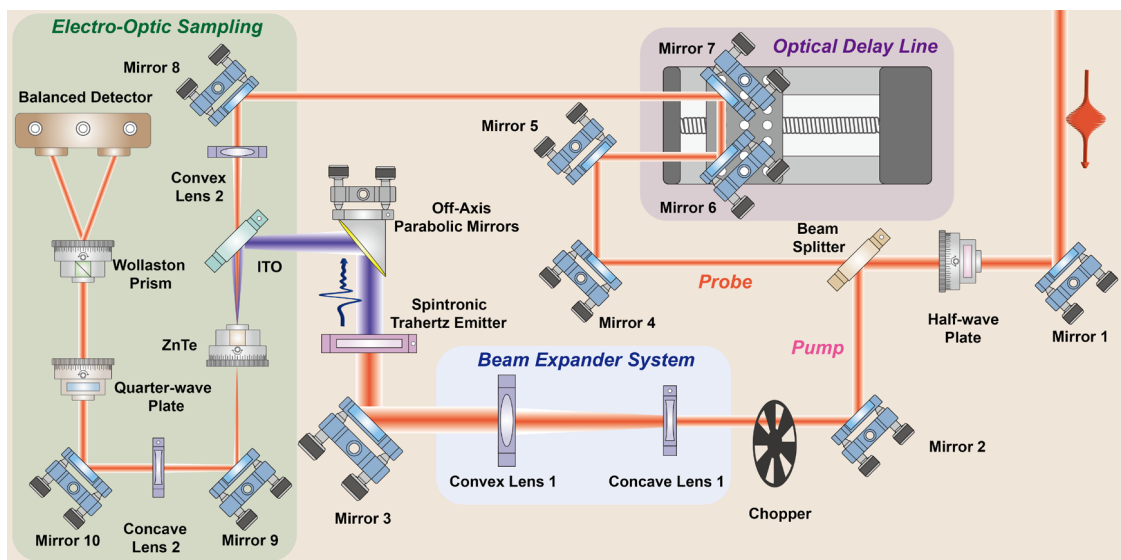

Figure 3. The schematic diagram of the THz emission spectroscopy setup

When the two samples had the same horizontal magnetic direction, 1.0 mJ energy was employed to pump each sample, and the time-domain signals obtained through electro-optical sampling are presented in Figure 4(a). The signal of the Si-PC-AFM-STEs is over 9 times that of the Si-AFM-STEs. Besides the nearly 2-fold increase in the efficiency of the heterostructure absorbing the pump pulse, the remaining enhancement mainly stems from the suppression of the photoconductive effect of the Si substrate, suggesting that a PC layer with a transmittance nearly 0 is necessary and highly effective for improving the performance of AFM-STEs. The above two time-domain signals were converted to frequency-domain signals in Figure 4(c) via Fourier transform. Compared with the sample with pure Si substrate, the sample with a PC layer shows an increase in amplitude and a spectrum covering more than 10 THz. Hence, it is an ideal ultra-broadband THz source, which is suitable for studying the interaction between THz and matter in the THz frequency window.

The incorporation of the PC layer maintains the dependence of the polarization of the $IrMn_3|Co_{20}Fe_{60}B_{20}|W$ heterostructure on the sample azimuth for THz radiation. The radiation pattern obtained by rotating the sample azimuth, as shown in Figure 4(c), still shows an "8" shaped symmetrical distribution. Consequently, we can continue to effortlessly manipulate the polarization of the THz radiation from the sample by adjusting the sample azimuth. Notably, the THz signal of Si-PC-AFM-STEs is approximately 2.25 times stronger than that of conventional AFM-STEs. This enhancement is attributed not only to the increased absorption efficiency of AFM-STEs by the PC layer but also to the impedance matching facilitated by the high refractive index of Si. By using the formula $2n_2/(n_2+n_{air})$, we can calculate the impact of different substrate refractive indices on the amplitude of the forward-emitted THz signal. The pump pulse of traditional AFM-STEs is incident from the substrate surface and exits from the heterostructure surface, so $n_2 = n_{air}$. However, Si-PC-AFM-STEs are incident from the heterostructure surface and exit from the substrate surface, where $n_2 = n_{Si}$. In this case, the amplitude of the forward-emitted THz is increased by 1.55 times compared to the previous one. Therefore, even if the THz pulse is absorbed by the Si substrate by approximately 30%, the THz amplitude of Si-PC-AFM-STEs can still be increased to more than twice that of traditional AFM-STEs. Overall, the PC layer with high reflectivity plays a vital role in isolating the pump laser between the Si and AFM-STEs, significantly reducing the influence of the photoconductive effect of the Si substrate, and simultaneously improving the absorption efficiency of the $IrMn_3|Co_{20}Fe_{60}B_{20}|W$ sample by reflecting the pump light to enhance the performance of THz radiation.

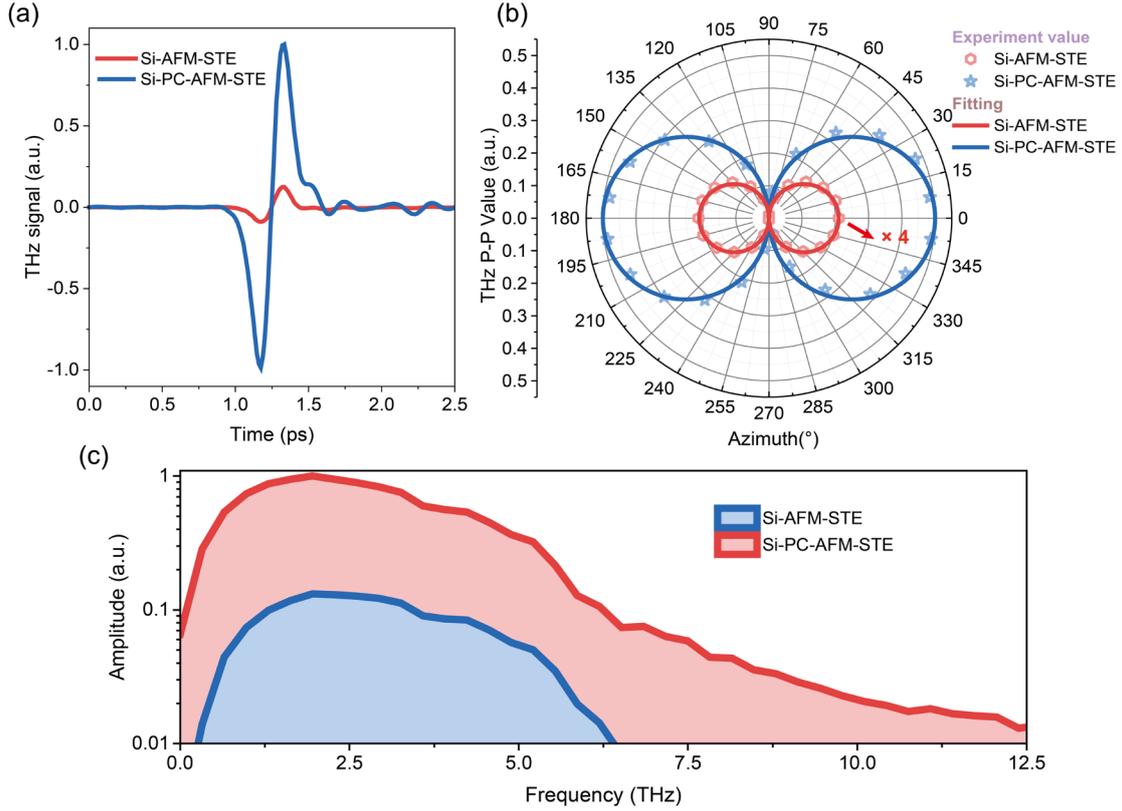

Figure 4. The effect of the PC layer on the performance of Si-AFM-STEs. (a) The THz time-domain signals of Si-AFM-STEs and Si-PC-AFM-STEs, (b) The radiation characteristics of rotating the sample azimuth, and (c) The THz frequency-domain signals.

To comparatively study the influence of substrate thermal conductivity on the AFM-STEs, $SiO_2$ was selected as the substrate because its thermal conductivity is significantly lower than that of Si (1/100 of Si). PC structures and AFM-STEs were deposited on the $SiO_2$ substrate by ion beam-assisted deposition and magnetron sputtering techniques respectively, to obtain $SiO_2$-PC-AFM-STEs samples. The Si-PC-AFM-STEs can reach a focal peak field up to 1.01 MV/cm with a single radiation energy of 62.5 nJ (Gentec SDX-1152) under the condition of 28 mm pump diameter and the pump energy of 5.5 mJ. However, under the same conditions, the peak field of $SiO_2$-PC-AFM-STEs is only 836 kV/cm, which is approximately 83% of that of the Si substrate. To further explore the effect of thermal effects on the heterostructure, a pump fluence dependence was carried out on the two samples with different substrates. The results shown in Figure 5(b) indicate that the electric field of both samples exhibits an exponential distribution as the pump fluence increases, reaching saturation at 0.65 mJ/cm$^2$. As the pump fluence increases, the heat deposition in the substrate becomes more and more obvious, leading to a larger gap, which is in accordance with the expected influence of substrate thermal conductivity. This relationship is further supported by the THz energy versus pump fluence, as depicted in Figure 5(c). The correlation between radiation energy and pump fluence follows a quadratic distribution. Due to the differences in substrate thermal conductivity, the difference in radiation energy between the two samples will gradually widen.

The pump pulse is incident perpendicularly onto the surface of the sample, forming a Fabry-Perot resonator between the parallel interfaces. After undergoing multiple reflections, the pulse exits from the interface to air. Although Si-PC-AFM-STEs can ignore the forward propagation of the pump pulse because of the PC layer, some energy is still lost due to the backward pump pulse. To solve this problem, a 70 nm thick MgO layer was added to the initial structure as an impedance matching layer. This layer has the double functions of reducing the backward pump laser and acting as a cap layer for $IrMn_3$ to prevent oxidation and extend the lifetime of the sample. Under the condition of a pump energy of 5.5 mJ, the MgO-coated sample showed an approximately 7.5% increase in THz electric field compared to the sample without the MgO layer, with the focal peak field reaching 1.085 MV/cm. This enhancement is mainly due to the absorption of the surplus backward pump pulse in the original structure by the heterostructure.

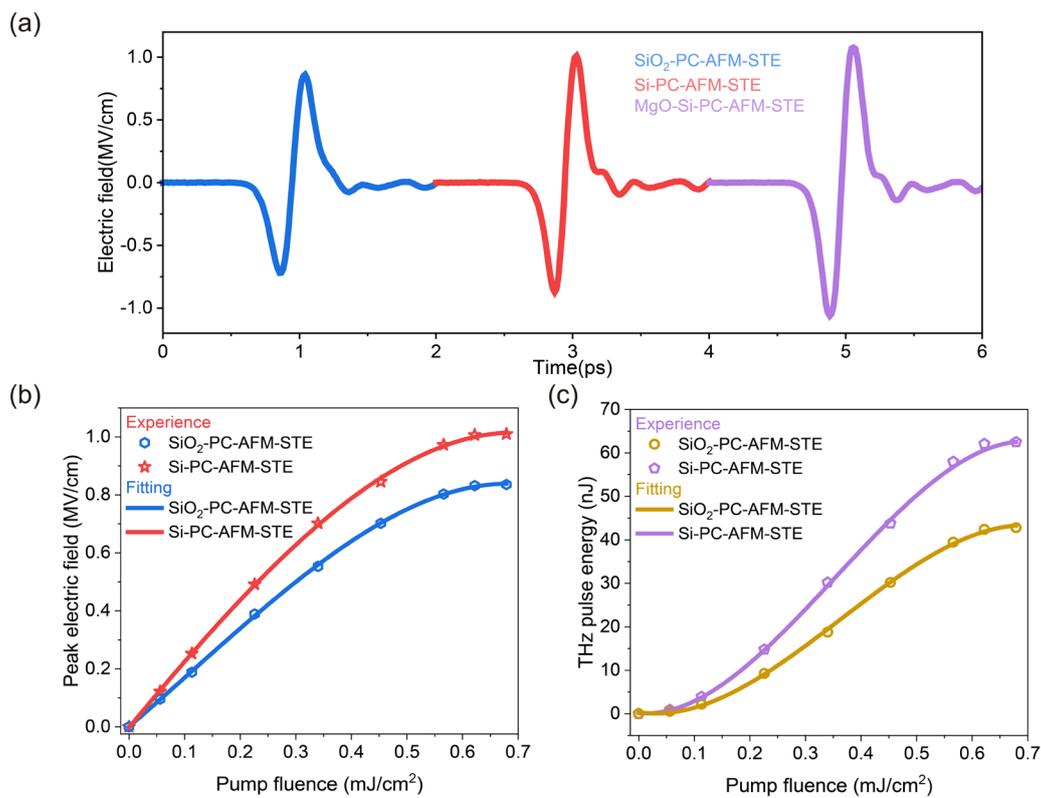

Figure 5. The influence of different substrates on the AFM-STEs. (a) The THz time-domain signals of $SiO_2$-PC-AFM-STEs, Si-PC-AFM-STEs, and MgO-Si-PC-AFM-STEs. (b) The pump fluence dependence of THz peak electric field. (c) The pump fluence dependence of THz energy.

The fabrication of STEs by magnetron sputtering provides the advantage of scalability in sample size. Additionally, it is not challenging to increase the pump light energy and pump beam. However, achieving large-area high-field THz radiation with materials such as W| $Co_{20}Fe_{60}B_{20}$| Pt encounters difficulties as the sample size expands. The main impediment lies in the hardship of applying a uniform external magnetic field that can saturate the magnetization of the FM layer. Experimental verification on a ~1-inch sample of W| $Co_{20}Fe_{60}B_{20}$| Pt, which comprises of a Si substrate, a PC layer, a heterostructure, and a MgO layer, demonstrated the influence of the external magnetic field. As depicted in Figures 6(a) and (b), a 6 mm beam was utilized to pump the sample, and the magnitude and direction of the external magnetic field were varied by moving the magnet along the x and y directions. To avoid the influence of the remanent magnetism in $Co_{20}Fe_{60}B_{20}$ caused by the external magnetic field on the test results, the magnets were moved from far to near. The experimental results are shown in Figure 6(c), where the horizontal and vertical coordinates respectively indicate the distances between the two magnets in the x and y directions. We use the color of the heat map to represent the changes in the electric field of the THz radiation. The study revealed that the amplitude of THz radiation increases as the distances decrease. This emphasizes the significance of precisely designing the external magnetic field structure to achieve uniform THz radiation over a large area for $W|Co_{20}Fe_{60}B_{20}|Pt$. The Halbach array has been proven to be capable of providing a uniform external magnetic field of 10 mT for 3-inch samples, but as the sample size further increases, it will be difficult for the samples to provide a suitable external magnetic field. Moreover, the existence of an external magnetic field may conflict with the advantages of miniaturization and integration of STEs.

To tackle the aforementioned problems, $IrMn_3|Co_{20}Fe_{60}B_{20}|W$, which functions independently of an external magnetic field, emerges as an optimal choice. As shown in Figure 6(d), a 5×5 array was selected on the surface of the 4-inch sample at equal intervals and scanned with a 6 mm pump beam in the absence of an external magnetic field. Figure 6(e) presents the emission of the sample at various positions, along with the corresponding horizontal and vertical coordinates of the sample positions. The normalized results reveal that the THz electric field within the 4-inch region maintains a high level above 0.95, a significant deviation from the emission results of $W|Co_{20}Fe_{60}B_{20}|Pt$. This difference is mainly attributed to the effective exchange bias or coupling effect at the $IrMn_3|Co_{20}Fe_{60}B_{20}$ interface, resulting in the existence of a uniform and stable in-plane magnetic field within the $Co_{20}Fe_{60}B_{20}$ layer.

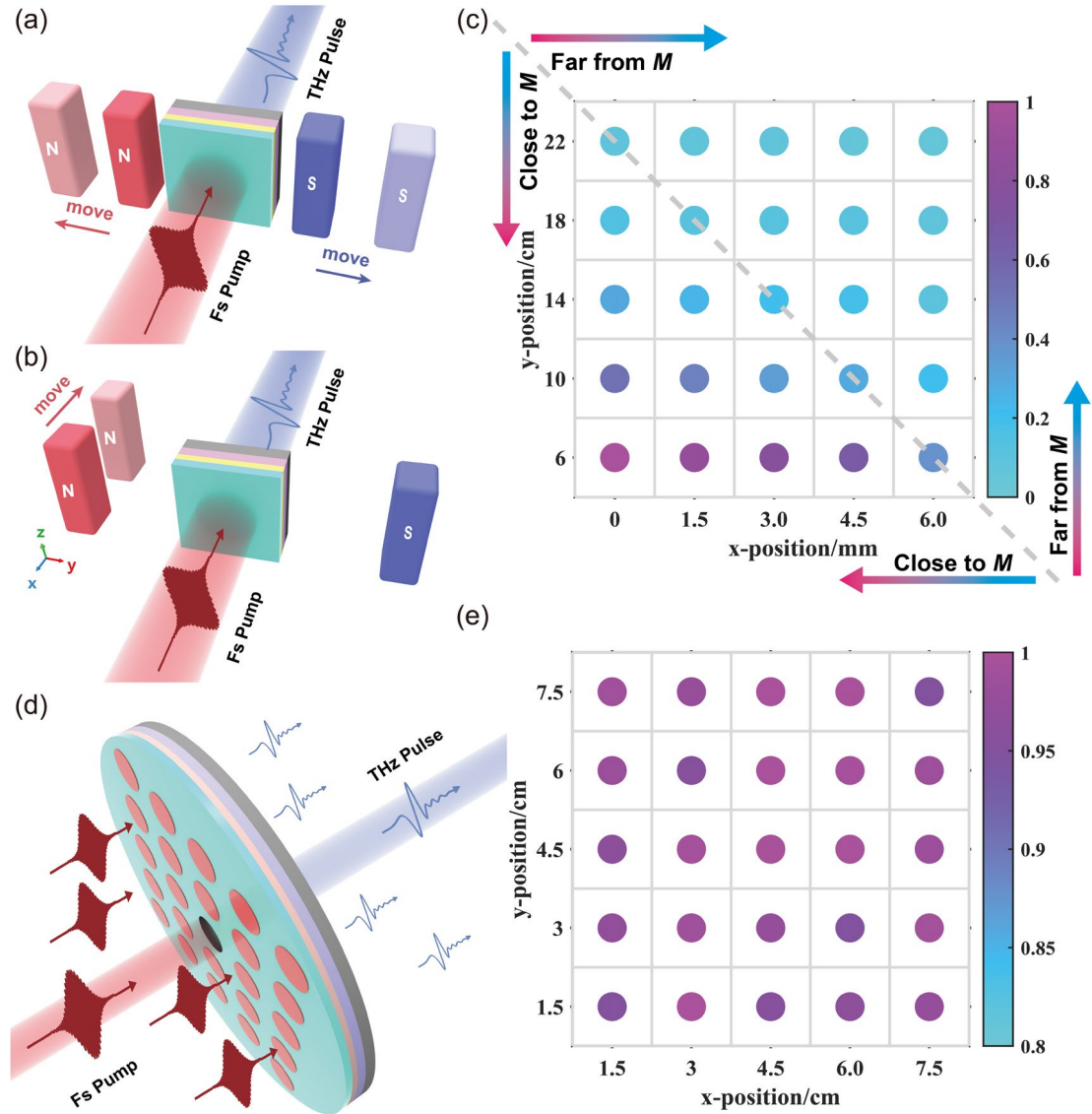

Figure 6. The THz radiation characteristics of W|Co$_{20}$Fe$_{60}$B$_{20}$|Pt and IrMn$_3$|Co$_{20}$Fe$_{60}$B$_{20}$|W. (a, b) The schematic diagrams of THz radiation of W|Co$_{20}$Fe$_{60}$B$_{20}$|Pt under different external magnetic fields and (c) The THz peak field strength. (d) The schematic diagram of THz radiation at different positions of the 4-inch IrMn$_3$|Co$_{20}$Fe$_{60}$B$_{20}$|W sample without an external magnetic field and (e) The THz peak field strength

## 3 Conclusion

Overall, the optimization of pump pulse utilization and the alleviation of the photoconductive effect of Si are accomplished by taking advantage of the high reflectivity of the one-dimensional photonic crystal layer [HfO$_2$(92 nm)|SiO$_2$(136 nm)]$_{40}$. This method effectively makes use of the superior thermal conductivity of the Si substrate and minimizes the influence of substrate heat deposition on the performance of the IrMn$_3$|Co$_{20}$Fe$_{60}$B$_{20}$|W heterostructure. The electric field of Si-PC-AFM-STEs is more than 2 times and 9 times higher than those of the SiO$_2$ substrate sample and the Si substrate sample respectively. With a 5.5 mJ energy pumping, the focal peak field of the sample can reach > 1.0 MV/cm. By making use of the exchange bias or coupling effect at the AFM|FM interface to pin the in-plane

magnetic field of the $Co_{20}Fe_{60}B_{20}$ layer and maintain a stable saturation magnetization within 4 inches, this approach reduces the significant impact of external magnetic fields on conventional STEs. It also enables the scaling of STEs sample sizes for larger areas of high-field THz radiation. This broadband high-field THz source, operating without an external magnetic field, holds potential for various applications, including the study of linear and nonlinear interactions between THz and matter. Furthermore, the horizontal propagation properties of the THz radiation suggest its potential as a crucial component for improving the performance of THz imaging systems.

4. Experimental Section

4.1 Sample processing

To obtain the desired samples in the study, it is essential to employ ion beam-assisted deposition and magnetron sputtering methods during the processing stage. Initially, a one-dimensional photonic crystal structure $[HfO2(92\ nm)|SiO2(136\ nm)]_{40}$ is fabricated on the surfaces of 4-inch double-sided polished Si and $SiO_2$ substrates via ion beam-assisted deposition to achieve high reflectivity within the 750-1050 nm range. After the initial deposition, the STE trilayer heterostructure is prepared by magnetron sputtering at room temperature. The fabrication of the $IrMn_3|Co_{20}Fe_{60}B_{20}|W$ heterostructure sample is carried out under the condition of a chamber pressure below $2\times10^{-8}$ Torr, with the induction of exchange bias by a 180 Oe in-plane external magnetic field to achieve saturation magnetization of the $Co_{20}Fe_{60}B_{20}$ layer. The $W|Co_{20}Fe_{60}B_{20}|Pt$ heterostructure needs to be prepared under a chamber pressure lower than $1\times10^{-7}$ Torr. It is crucial to note that the sequence of coating in the trilayer heterostructure involves the deposition of the W layer on the substrate first, followed by the sequential deposition of $Co_{20}Fe_{60}B_{20}$ and Pt ($IrMn_3$).

4.2 THz emission spectroscopy

Figure 3 presents a schematic illustration of the experimental setup. A Ti: sapphire laser amplifier with central wavelength of 800 nm, pulse duration of 35 fs, and repetition frequency of 1 kHz was employed as the source. The beam was divided into pump pulse and probe pulse using a 9:1 beam splitter. Before irradiating the sample with the pump laser, the collimated pump beam was expanded to 2 cm (half-peak width of the intensity) by a 3× expansion system composed of a concave lens (-50 mm) and a convex lens (150 mm). The THz radiation from the sample was guided through an off-axis parabolic mirror with a focal length of 4-inches and an ITO, and then focused onto a 0.1 mm (110)-oriented ZnTe detection crystal. The detection optical setup mainly consisted of optical delay line module and standard electro-optical sampling module. The probe laser was focused onto the surface of the detection crystal via a convex lens, creating a spatial combination with the focused THz. Subsequently, after being collimated by the concave lens, the pulse traveled parallel to the subsequent electro-optical sampling module for the acquisition of THz time-domain signals.